\documentclass[12pt]{article}
\usepackage{graphicx}
\usepackage{subcaption}
\usepackage{lineno}

\newcommand\blfootnote[1]{%
  \begingroup
  \renewcommand\thefootnote{}\footnote{#1}%
  \addtocounter{footnote}{-1}%
  \endgroup
}

\newcommand\pubnumber{SNSN-323-63}
\newcommand\pubdate{\today}

\def\institute{Faculty of Mathematics, Physics and Informatics\\ 
Comenius University in Bratislava, Mlynsk\'a dolina 84248 Bratislava, Slovakia}
\def\support{\blfootnote{\hspace{-0.75cm} \copyright~2019 CERN for the benefit of the ATLAS Collaboration. \\ Reproduction of this article or parts of it is allowed as specified in the CC-BY-4.0 license.}}

\def\Title#1{\begin{center} {\Large #1 } \end{center}}
\def\Author#1{\begin{center}{ \sc #1} \end{center}}
\def\Address#1{\begin{center}{ \it #1} \end{center}}

\newcommand\pubblock{\rightline{\begin{tabular}{l} \pubnumber\\
         \pubdate  \end{tabular}}}
\newenvironment{Abstract}{\begin{quotation}  }{\end{quotation}}
\newenvironment{Presented}{\begin{quotation} \begin{center} 
             PRESENTED AT\end{center}\bigskip 
      \begin{center}\begin{large}}{\end{large}\end{center} \end{quotation}}

\def\beq{\begin{equation}}
\def\eeq#1{\label{#1}\end{equation}}
\def\eeqn{\end{equation}}

\def\beqa{\begin{eqnarray}}
\def\eeqa#1{\label{#1}\end{eqnarray}}
\def\eeqan{\end{eqnarray}}

\let\bar=\overbar

\def\Dslash{\not{\hbox{\kern-4pt $D$}}}
\def\dslash{\not{\hbox{\kern-2pt $\del$}}}

\def\msb{{\bar{\ssstyle M \kern -1pt S}}}

\begin{document}
\begin{titlepage}
\pubblock

\vfill
\Title{Top-antitop charge asymmetry measurements in the lepton+jets channel with the ATLAS detector}
\vfill
\Author{ Matej Melo, on behalf of the ATLAS Collaboration \support}
\Address{\institute}
\vfill
\begin{Abstract}
We report a measurement of the charge asymmetry in top quark pair production with the ATLAS experiment at the LHC using 20.3 fb$^{-1}$ of $pp$ collision data at $\sqrt{s} = 8$ TeV. The data are unfolded using a fully Bayesian unfolding method. Differential measurements of the charge asymmetry are performed.
\end{Abstract}
\vfill
\begin{Presented}
$11^\mathrm{th}$ International Workshop on Top Quark Physics\\
Bad Neuenahr, Germany, September 16--21, 2018
\end{Presented}
\vfill
\end{titlepage}
\def\thefootnote{\fnsymbol{footnote}}
\setcounter{footnote}{0}

\section{Introduction}
	Charge asymmetry $A_{\mathrm{C}}$ in top-antitop quark pair ($t\bar t$) production is defined as:
	$$A_{\mathrm{C}}^{t\bar t}=\frac{N(\Delta |y| > 0) - N(\Delta |y| < 0)}{N(\Delta |y| > 0) + N(\Delta |y| < 0)},$$
	where $\Delta |y| = |y_t|-|y_{\bar t}|$ and $y_t$ ($y_{\bar t}$) is the rapidity of the top (antitop) quark.  
	\begin{figure}[htb]
		\begin{center}
			~~~~~~~~~\includegraphics[width=0.23\textwidth]{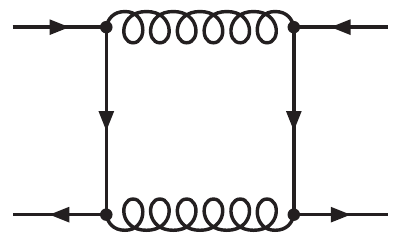}	
		~~~~~~
			\includegraphics[width=0.23\textwidth]{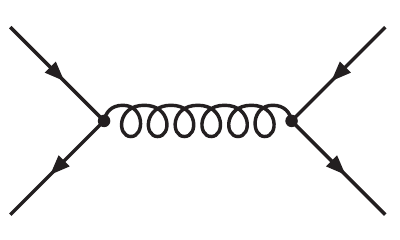}~~~~~~~~~~~
		\caption{Box (left) and Born (right) diagrams \cite{Kuhn:2011ri}.}
		\label{boxBorn}
		\end{center}
	\end{figure}
	In the Standard Model (SM), non-zero asymmetry is predicted due to interference of higher order quark-antiquark annihilation diagrams. 
Main contribution to the $A_{\mathrm{C}}$ comes from the Box and Born diagram interference  \cite{Kuhn:2011ri}, shown in Figure \ref{boxBorn}. 
Many theories beyond the Standard Model (BSM) predict an enhancement of the asymmetry, especially in the region of high $t\bar t$ invariant mass. 
Two measurements were performed by the ATLAS experiment \cite{ATLAS} at the LHC using 20.3 fb$^{-1}$ of $pp$ collision data at \mbox{$\sqrt{s} = 8$ TeV} in the lepton+jets channel: 
in the so-called resolved topology \cite{ATLASResolved} and using events with highly boosted top or antitop quarks \cite{ATLASBoosted}.


\section{Event Selection and Reconstruction}
	Single lepton trigger requirements are applied in the lepton+jets decay channel and exactly one good lepton (electron or muon) with $p_{\mathrm{T}} > 25$ GeV is required in both topologies. 
To suppress background contribution various criteria on missing transverse energy $E_{\mathrm{T}}^{\mathrm{miss}}$ are applied, depending on the multiplicity of $b$-tagged jets. 
In resolved topology, at least four jets with $p_{\mathrm{T}} > 25$ GeV are required. 
Kinematic Likelihood Fitter \cite{KLFitter} is then used to reconstruct $t\bar t$ kinematics and events are separated into six signal regions based on the lepton charge ($+1,-1$) and $b$-tag multiplicity (0, 1, $\geq 2$), obtained by using a working point of the $b$-tagging algorithm corresponding to an efficiency of 70\% for identifying $b$-jets.  
In the boosted topology, at least one jet with $p_{\mathrm{T}} > 25$ GeV close to the charged lepton ($\Delta R < 1.5$) is required. The boosted topology requires also at least one large radius jet ($R=1.0$) with $p_{\mathrm{T}} > 300$ GeV, top-quark tagged by using jet substructure variables and well separated from the lepton with $\Delta \phi(\ell,\mathrm{jet}_{R = 1.0}) > 2.3$. Either the jet close to the lepton or the jet matched to the large jet within $\Delta R < 1.5$ must be $b$-tagged using the 70\% efficient working point.

\section{Unfolding}
\begin{figure}[htb]
		\centering
		\includegraphics[width=0.6\textwidth,trim={0 0.5cm 0 0},clip]{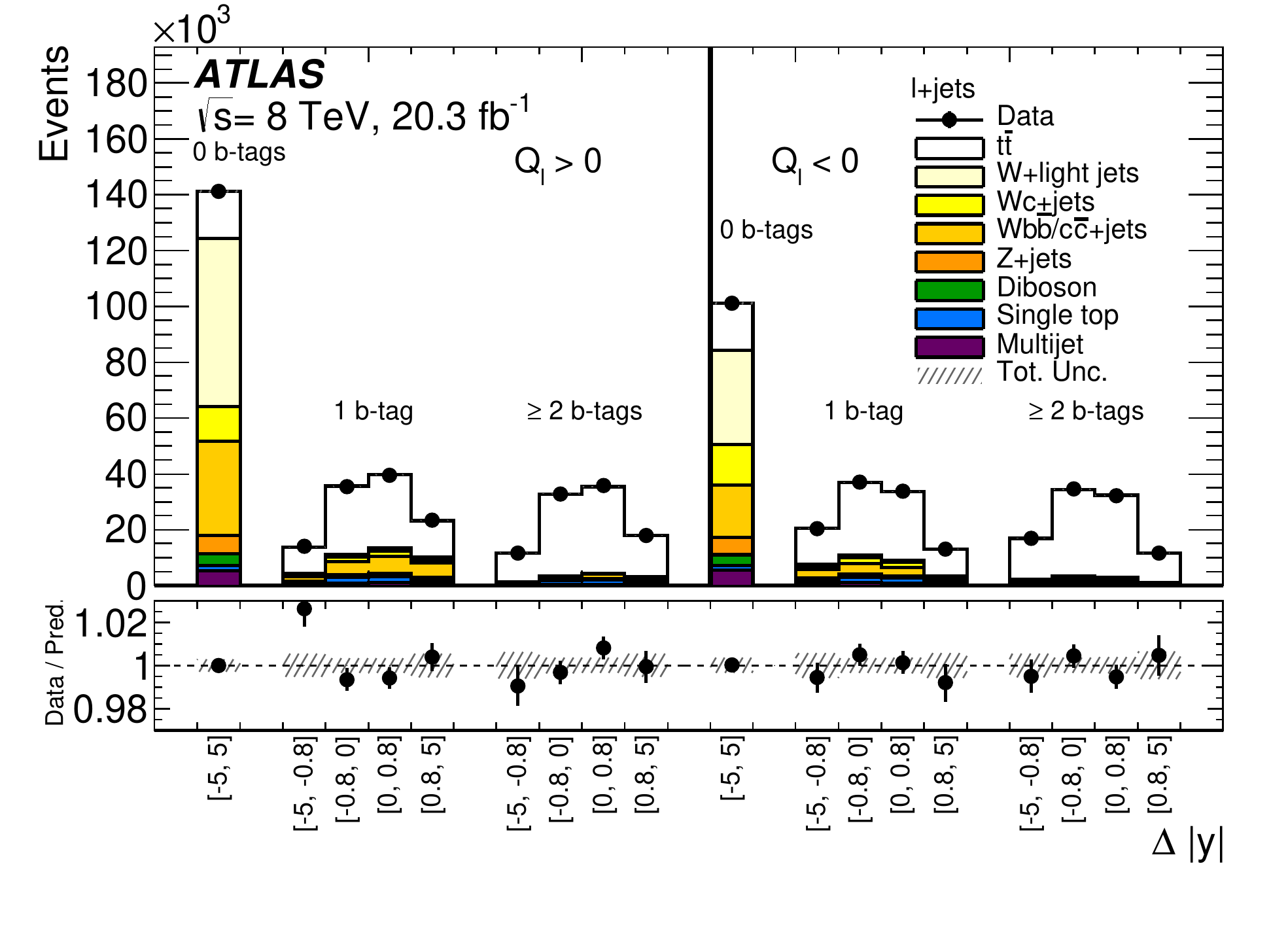}	
		\caption{Comparison between prediction and data for the 18 bins used in the inclusive $A_{\mathrm{C}}$ measurement in the resolved topology after applying the simultaneous unfolding procedure and $W$+jets in situ calibration \cite{ATLASResolved}.}
		\label{postfit}
	\end{figure}

Fully Bayesian Unfolding (FBU) \cite{FBU} is used in both measurements to unfold the reconstructed distributions to the parton level. 
In the case of the boosted topology, the measurement is performed in a fiducial phase space ($m_{t\bar t} > 0.75$ TeV and \mbox{$-2 < \Delta |y|< 2$}) due to small acceptance outside this region. 
For all systematic uncertainties, except those related to Monte Carlo generator modelling and the unfolding procedure itself, nuisance parameters are assigned. 
FBU enables to marginalize and to contrain systematic uncertainties. In the resolved topology \mbox{0-$b$ tag} region is used for in situ calibration of the $W$+jets background, Figure \ref{postfit}.


\section{Results}
\begin{figure}[htb]
			\centering
			\includegraphics[height=4.5cm,trim={0 0 0 0},clip]{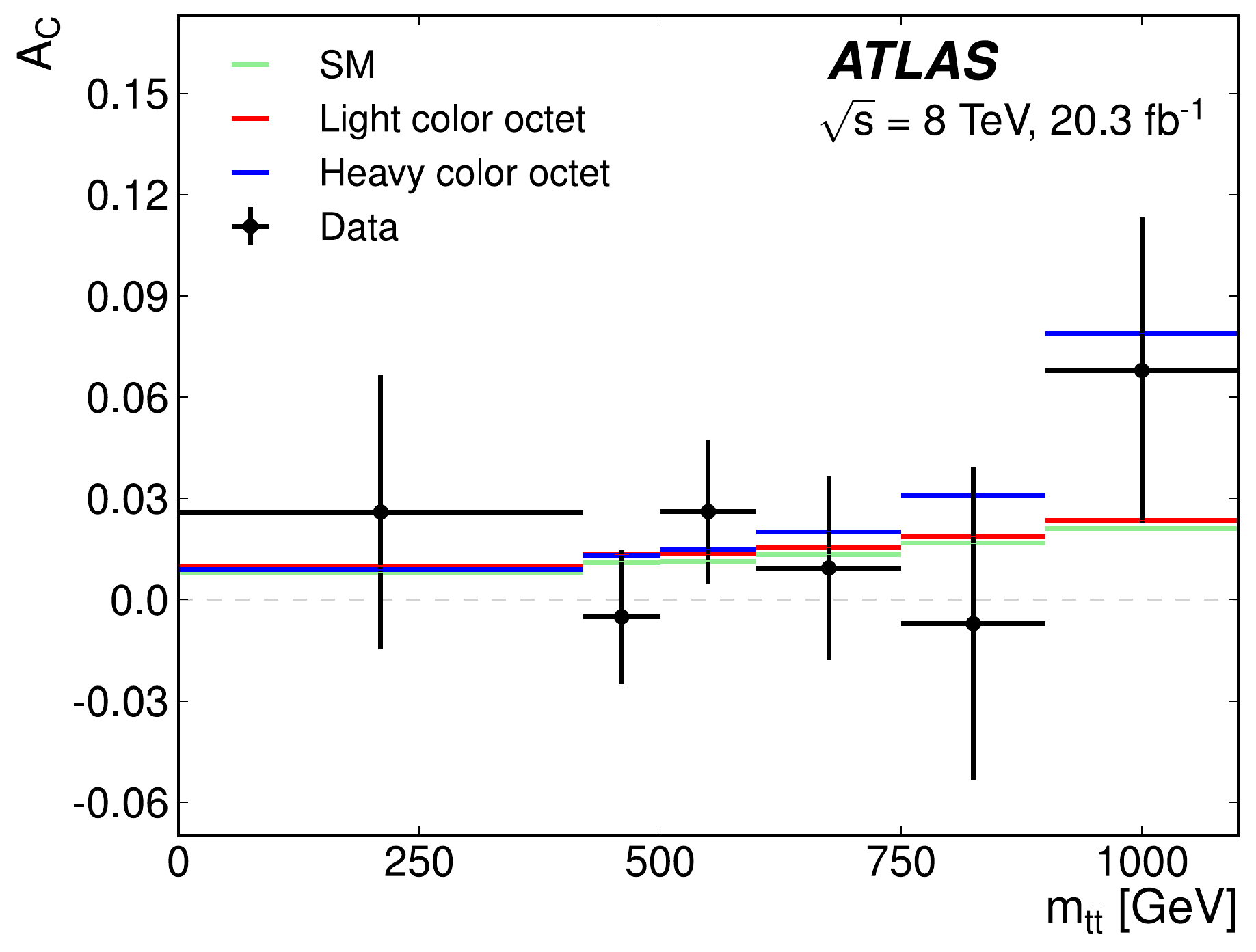} 
		~~
			\centering
			\includegraphics[height=4.5cm]{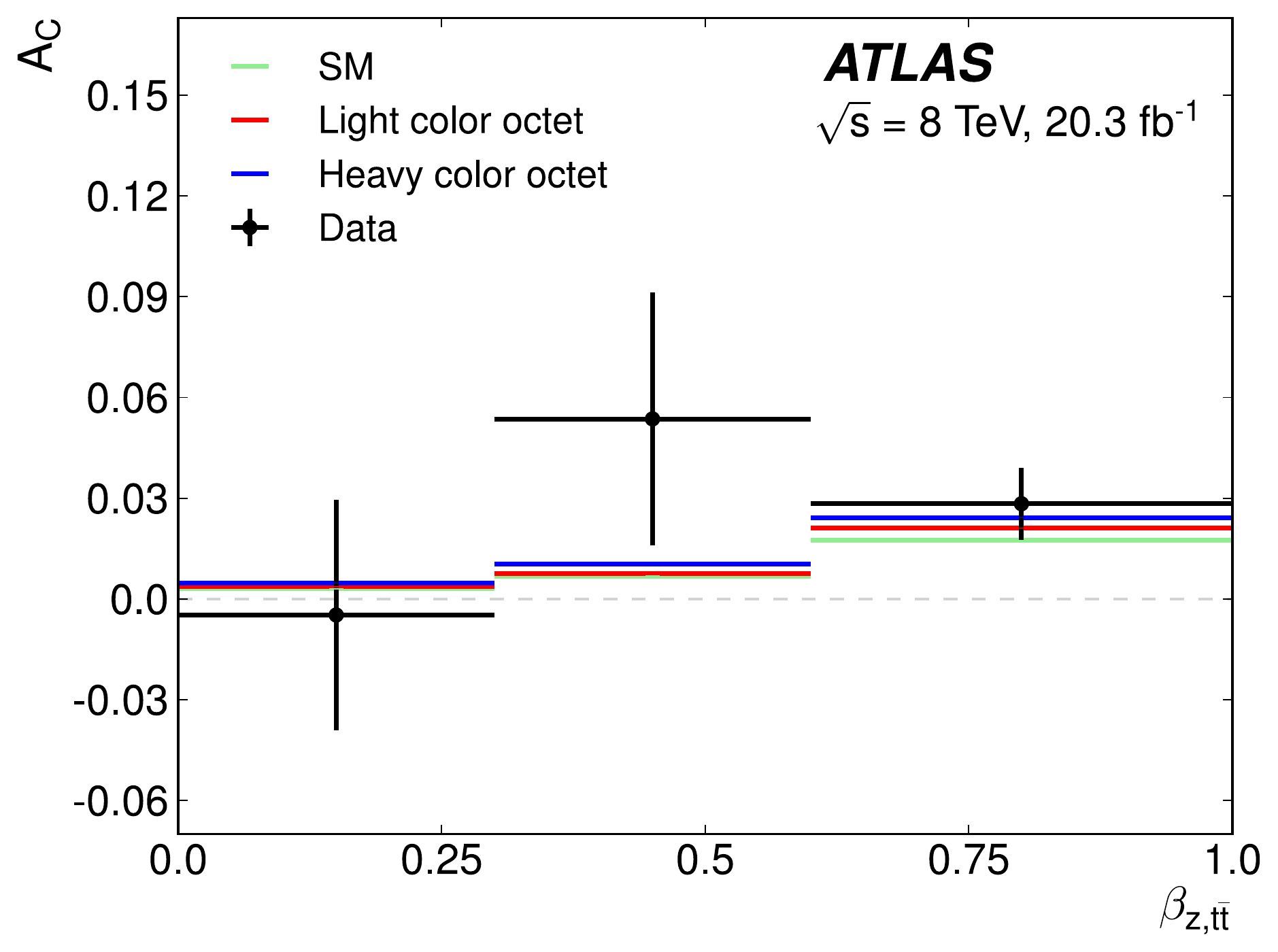} \\
			\vspace{0.3cm}
			\centering
			\includegraphics[height=4.5cm]{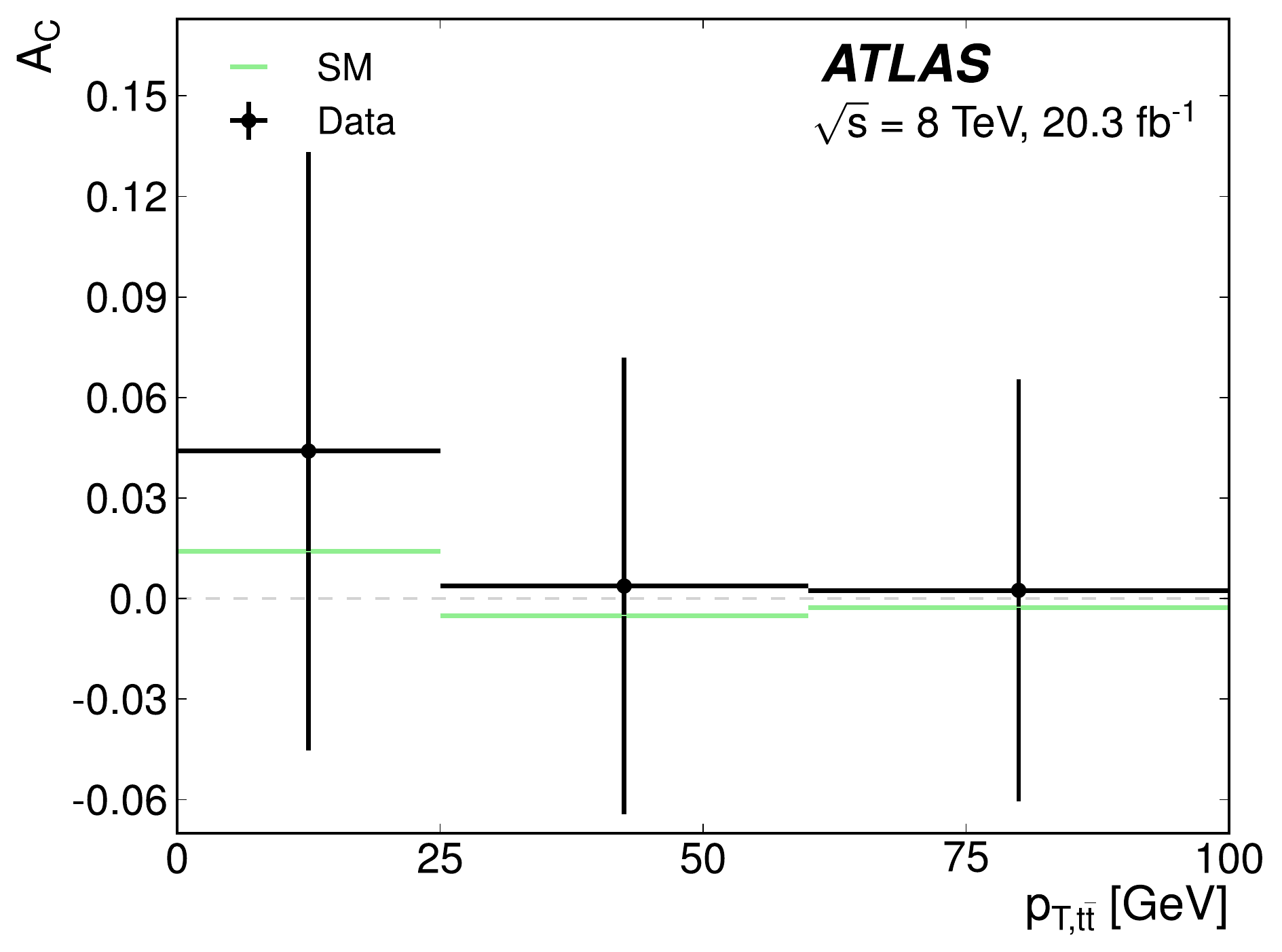} ~~~
		~~~~
			\centering
			\includegraphics[height=4.5cm,trim={0 0.4cm 0 0},clip]{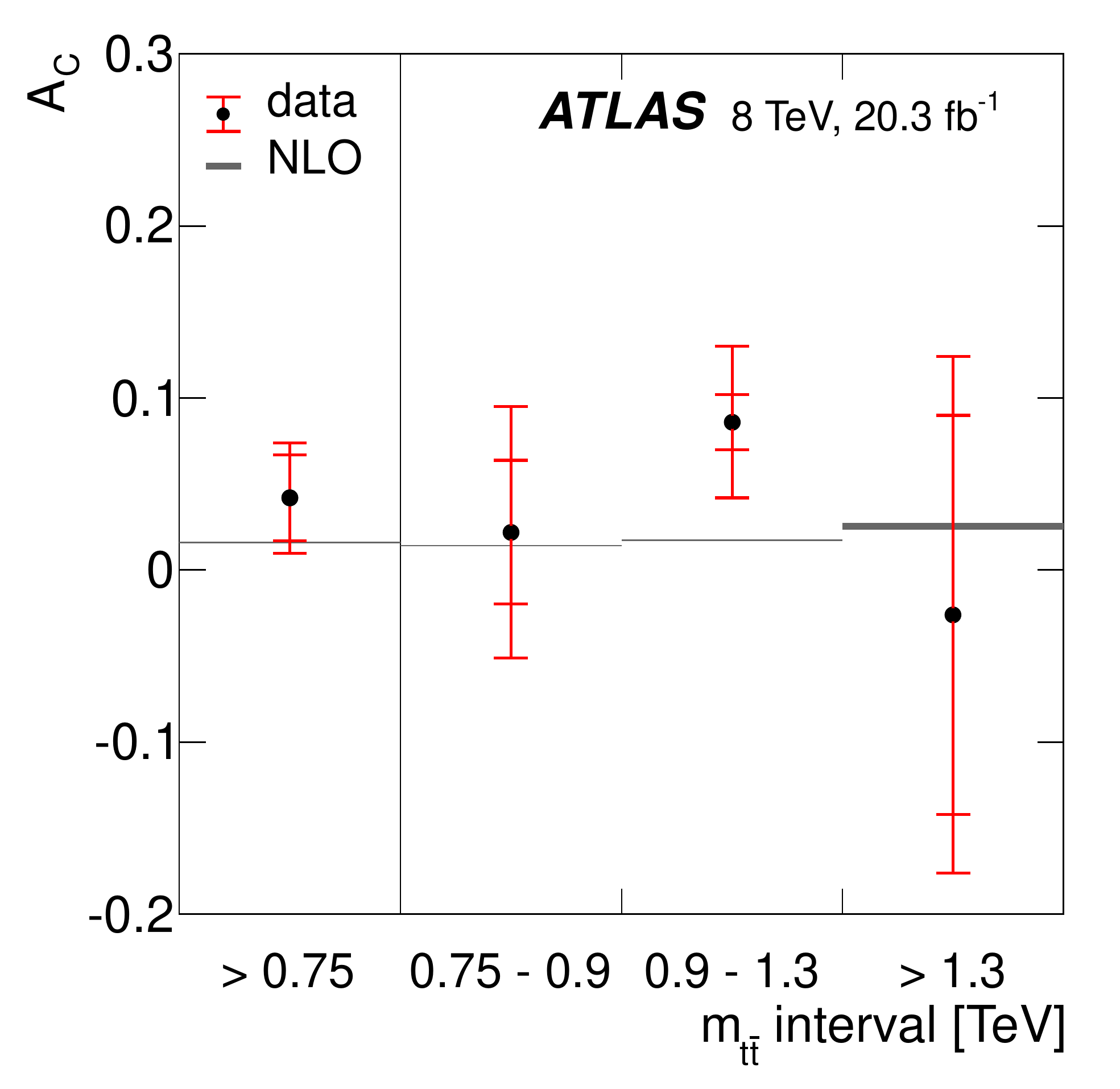}~~~
		\caption{Measured $A_{\mathrm{C}}$ values in the resolved topology as a function of bin-averaged $t\bar t$ mass (top left), $\beta_z$ (top right), $p_{\mathrm{T}}$ (bottom left) and fiducial and mass-dependent measurement in the boosted topology (bottom right). Measured values are compared with the SM predictions and in the case of the upper two plots also with BSM predictions \cite{ATLASResolved, ATLASBoosted}.}
		\label{results}
	\end{figure}

	The inclusive asymmetry measured in the resolved topology is $A_{\mathrm{C}} = 0.009\pm0.005$ (stat. + syst.), compatible with the Standard Model prediction $A_{\mathrm{C}}^{\mathrm{SM}} = 0.0111\pm0.0004$ \cite{Bernreuther:2012sx}. 
The fiducial asymmetry ($m_{t\bar t} > 0.75$ TeV and $-2 < \Delta |y|< 2$) measured in the boosted topology is $A_{\mathrm{C}} = 0.042\pm0.032$, compatible with $A_{\mathrm{C}}^{\mathrm{SM}} = 0.0160\pm0.0004$ \cite{Kuhn:2011ri}. 
Results of the $A_{\mathrm{C}}$ measurements in the resolved topology as a function of mass, $\beta_z$ and $p_{\mathrm{T}}$ of the $t\bar t$ system are shown in Figure \ref{results}, as well as the mass-dependent measurement in the boosted topology. 
Inclusive and differential measurements as a function of the $t\bar t$ mass are mostly limited by statistical uncertainties; measurements as a function of the top pair $\beta_z$ and $p_{\mathrm{T}}$ are mostly limited by modelling uncertainties. No significant deviations from the SM predictions are observed.

\subsection{Impact on BSM Scenarios}
\begin{figure}
		\begin{center}
		\includegraphics[width=0.4\textwidth]{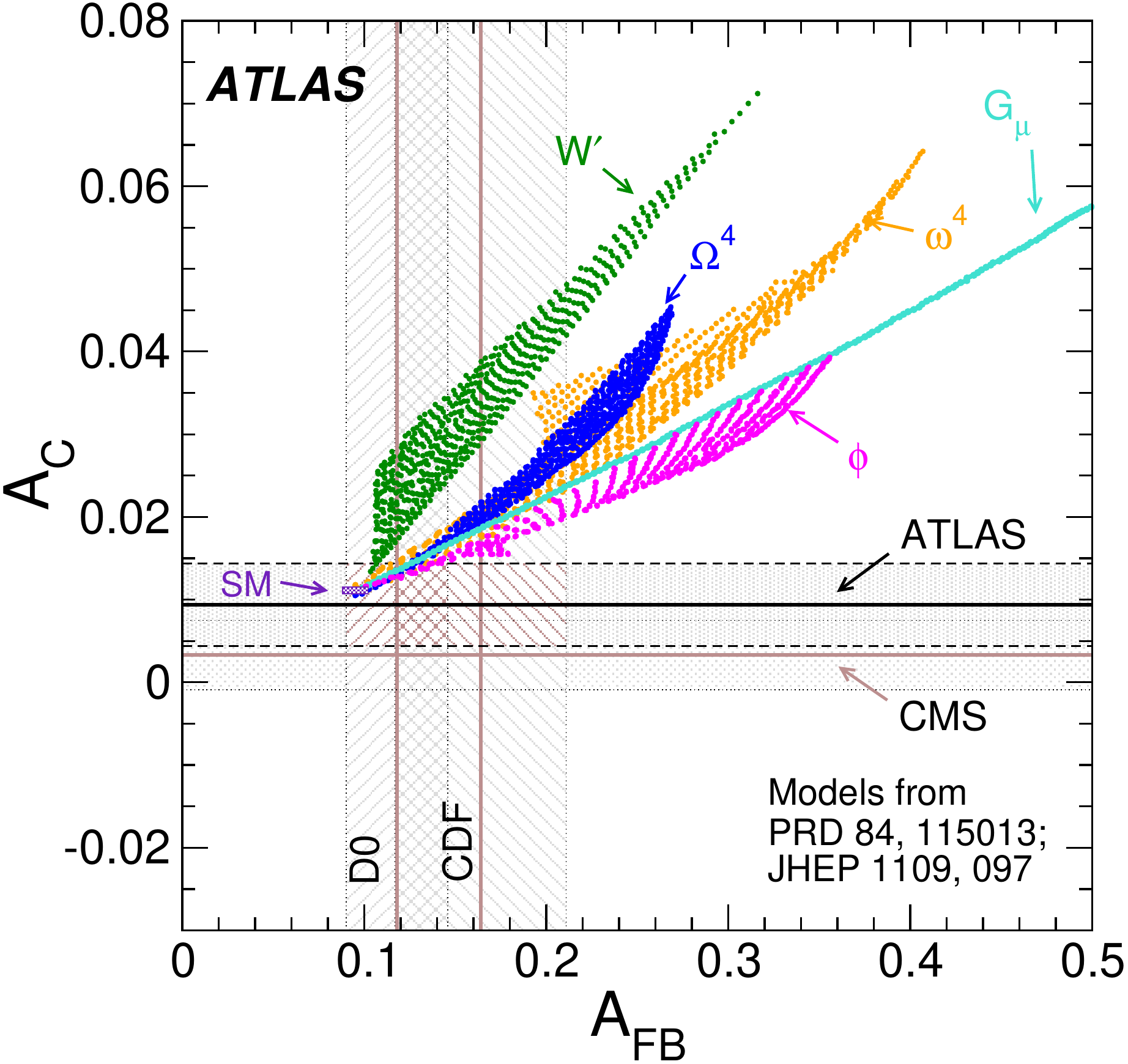} ~~~~~
		\includegraphics[width=0.4\textwidth]{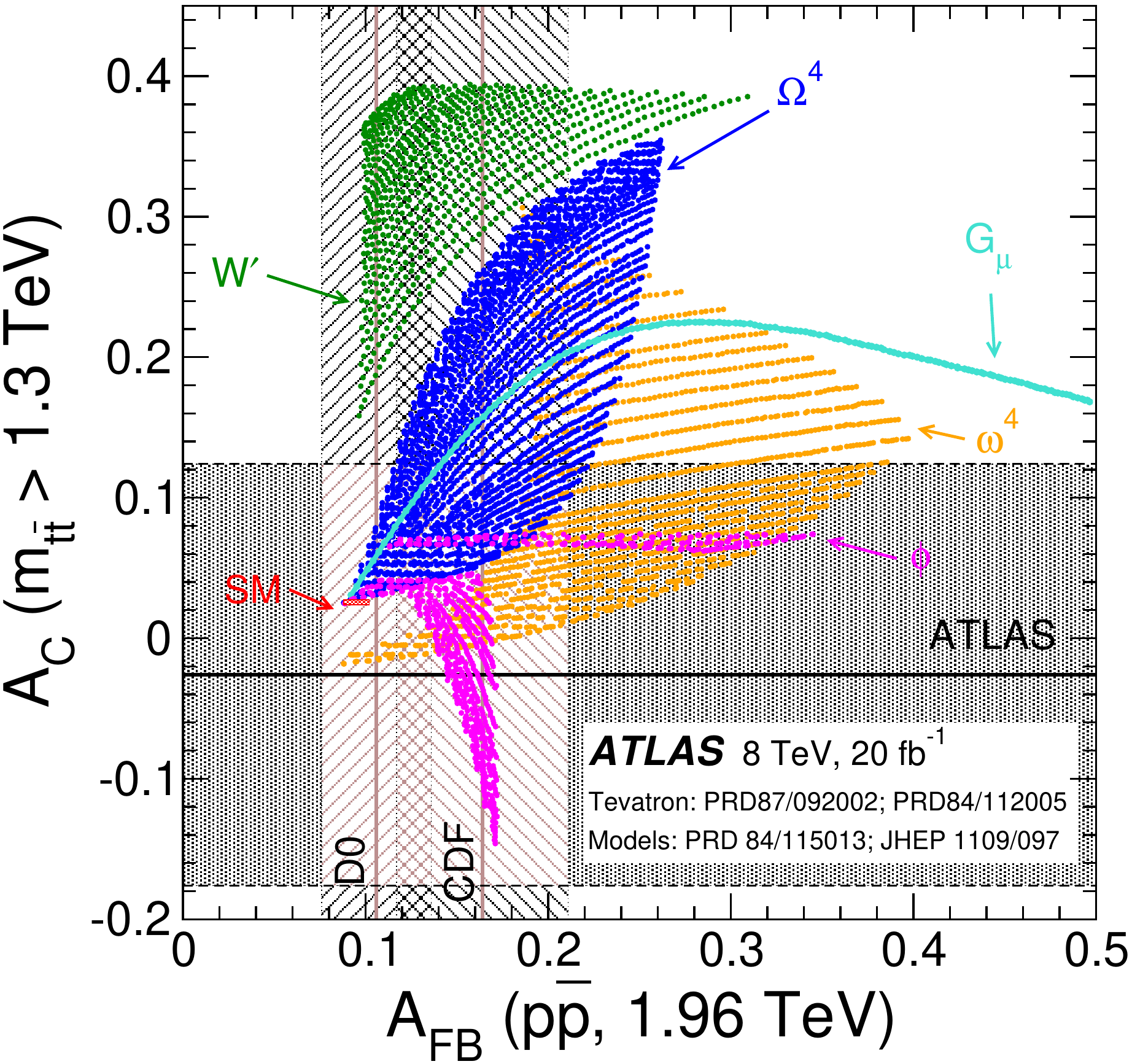}\\
		\end{center}
		\caption{Measured $A_{\mathrm{C}}$ inclusive (left) and in the high mass region of $m_{t\bar t} > 1.3$ TeV (right) versus the Tevatron forward-backward asymmetry $A_{\mathrm{FB}}$ results, compared with the SM predictions as well as predictions incorporating various potential BSM contributions \cite{BSM1,BSM2}. The uncertainty bands correspond to 68\% confidence level interval \cite{ATLASResolved,ATLASBoosted}.}
		\label{BSM}
	\end{figure}

	The ATLAS 8 TeV $A_{\mathrm{C}}$ measurements in the lepton+jets channel allow for the exclusion of a large region in the parameter space of various BSM models, as demonstrated in Figure \ref{BSM}. The cloud of points in Figure \ref{BSM} corresponds to a number of models described in Refs. \cite{BSM1, BSM2}: a $W'$ boson, a heavy axigluon $G_\mu$, a scalar isodoublet $\phi$, a colour-triplet scalar $\omega^4$ and a colour-sextet scalar $\Omega^4$. Each point corresponds to a choice of the new particle's mass in the range between 100 GeV and 10 TeV, as well as a choice of the couplings to SM particles which are required to be compatible with the Tevatron cross-section measurements. 

\section{Summary}
Measurements of the $t\bar t$ charge asymmetry $A_{\mathrm{C}}$ by the ATLAS experiment at the LHC using 20.3 fb$^{-1}$ of $pp$ collision data in the lepton+jets topology are consistent with the SM predictions and disfavour a large phase-space of parameters describing several BSM scenarios.


\end{document}